\documentclass{aa}
\usepackage[T1]{fontenc}
\usepackage[utf8]{inputenc}

\usepackage{newtxtext}  
\usepackage[varg]{newtxmath}  
\usepackage{graphicx}
    \DeclareGraphicsExtensions{.pdf}
\usepackage{float}
\usepackage{amsmath}    
\usepackage{amssymb}    
\usepackage{marvosym}   
\usepackage{nicefrac}   
\usepackage{textcomp}
\usepackage{url}
\usepackage{bm}         
\usepackage{tikz}
\usepackage{tikzscale}
\usepackage{mathtools}
\usepackage{afterpage}
\usepackage[exponent-product=\cdot
            ,tight-spacing=true
            ,range-phrase={\text{--}}
            ,range-units=single
            ,list-units=single
            ,angle-symbol-over-decimal=true
            ,detect-weight=true
            ]{siunitx}
\DeclareSIUnit\au{au}
\DeclareSIUnit\pc{pc}
\DeclareSIUnit\jy{Jy}
\DeclareSIUnit\msun{M\ensuremath{_{\sun}}}
\DeclareSIUnit\mjup{M\ensuremath{_{\text{\Jupiter}}}}
\DeclareSIUnit\lsun{L\ensuremath{_{\sun}}}
\DeclareSIUnit\Mlambda{\si{\mega}\ensuremath{\lambda}}

\usepackage{hyperref}
\usepackage{prettyref}  
\newrefformat{fig}{\text{Fig.~\ref{#1}}}
\newrefformat{eq}{\text{Eq.~\ref{#1}}}
\newrefformat{tab}{\text{Table~\ref{#1}}}
\newrefformat{sec}{\text{Sect.~\ref{#1}}}
\newrefformat{enu}{\text{~\ref{#1}}}

\newcommand{\Tc}        {\ensuremath{T_{\text{c}}}}
\newcommand{\Lc}        {\ensuremath{L_{\text{c}}}}
\newcommand{\bfcal}[1]  {\ensuremath{\bm{\mathcal{#1}}}}

\title{Constraints on observing brightness asymmetries in protoplanetary disks at solar system scale}
\titlerunning{Brightness asymmetries at solar system scale}

\author{Robert Brunngräber \and Sebastian Wolf}
\institute{Institut für Theoretische Physik und Astrophysik, Christian-Albrechts-Universität zu Kiel, Leibnizstr. 15, 24118 Kiel, Germany\\\email{rbrunngraeber@astrophysik.uni-kiel.de}}

\date{Received / Accepted}

\abstract{We have quantified the potential capabilities of detecting local brightness asymmetries in circumstellar disks with the Very Large Telescope Interferometer (VLTI) in the mid-infrared wavelength range. The study is motivated by the need to evaluate theoretical models of planet formation by direct observations of protoplanets at early evolutionary stages, when they are still embedded in their host disk. Up to now, only a few embedded candidate protoplanets have been detected with semi-major axes of 20--50\,au. Due to the small angular separation from their central star, only long-baseline interferometry provides the angular resolving power to detect disk asymmetries associated to protoplanets at solar system scales in nearby star-forming regions. In particular, infrared observations are crucial to observe scattered stellar radiation and thermal re-emission in the vicinity of embedded companions directly. For this purpose we performed radiative transfer simulations to calculate the thermal re-emission and scattered stellar flux from a protoplanetary disk hosting an embedded companion. Based on that, visibilities and closure phases are calculated to simulate observations with the future beam combiner MATISSE, operating at the \textit{L}, \textit{M} and \textit{N} bands at the VLTI. We find that the flux ratio of the embedded source to the central star can be as low as 0.5 to 0.6\,\% for a detection at a feasible significance level due to the heated dust in the vicinity of the embedded source. Furthermore, we find that the likelihood for detection is highest for sources at intermediate distances $r\approx2$--$5$\,au and disk masses not higher than $\approx10^{-4}$\,M$_{\odot}$.}

\bibpunct{(}{)}{;}{a}{}{,}

\begin{document}
\maketitle



\section{Introduction}
    Based on theoretical models, planets are believed to form inside dense, gas-rich protoplanetary disks, and the presence of complex, selected large-scale structures that have been observed for a large number of protoplanetary disks are thought to be linked to planet-disk interaction \citep[e.g.][]{baruteau-et-al-2014}. Several scenarios for the formation and evolution of planets are discussed in the literature. For jovian planets, the two competing, most commonly used theories are known as the `cold' and `hot start' scenarios, corresponding to the formation and evolution via core accretion and gravitational instability, respectively \citep{pollack-et-al-1996,boss-1997,burrows-et-al-1997,marley-et-al-2007}. The luminosity of forming -- and thus embedded -- planets can be used to approximate their entropy, which has a diagnostic character for the two formation processes, although it is not unambiguous \citep{mordasini-et-al-2012,mordasini-2013,szulagyi-mordasini-2017}. Therefore, the detection and analysis of protoplanets is essential to verify selected stages of the theoretically predicted planet formation scenarios.
    
    Recently, first candidate protoplanets have been found via high-contrast imaging (HD~100546b: \citealt{quanz-et-al-2013,currie-et-al-2014,quanz-et-al-2015}; HD~169142b: \citealt{reggiani-et-al-2014}) and the non-redundant aperture masking technique (LkCa~15b,c: \citealt{sallum-et-al-2015}). These were the first direct observations of potential planets in a very early phase of their evolution, still embedded in their host disk. However, the spatial resolution of these observations are limited to $\approx \SIrange{20}{50}{\au}$ at distances of \SI{140}{\pc}, where a large fraction of the best studied protoplanetary disks is located in, for example, the \textit{Taurus-Aurigae} and \textit{$\rho$-Ophiuchi} star forming regions \citep{torres-et-al-2007,mamajek-2008}. With near and mid infrared (NIR, MIR), long-baseline interferometry the spatial resolution is increased to (sub-)au-scales, and observations are thus suitable to study planet formation in the habitable zone where terrestrial planets are expected to form.
    
    Using data obtained with MIDI (MID-infrared Interferometric instrument: \citealt{leinert-et-al-2003a,leinert-et-al-2003b,morel-et-al-2004}) at the Very Large Telescope Interferometer (VLTI) at three different epochs, \citet{brunngraeber-et-al-2016} found evidence for temporal brightness variations as well as non-rotationally symmetric structures in the protoplanetary disk around DR~Tau at stellar separations as small as $1$ -- $4$\,au. In that study, we could prove that hotspots, potentially generated by protoplanet-disk interactions, can be identified with a two-beam MIR interferometer. The upcoming second generation instrument MATISSE (Multi-AperTure mid-Infrared SpectroScopic Experiment) for the VLTI \citep{lopez-et-al-2014} will allow users to continue milli-arcsecond observations in the MIR (\textit{N} band: \SIrange{8}{13}{\um}). In addition, observations in the \textit{L} (\SIrange{3}{4}{\um}) and \textit{M} (\SIrange{4.5}{5}{\um}) bands will become possible. With the combination of four beams, MATISSE will be able to use six baselines simultaneously, opening the ability to include the analysis of closure phases in the MIR.
    
    In this paper, we quantify the ability of MATISSE to detect disk asymmetries -- hot spots -- potentially caused by embedded protoplanets at solar-system scales. A description of these hot spot asymmetries and the set-up of the underlying disk model is given in \prettyref{sec:setup_asy} and \prettyref{sec:setup_disk}, respectively. In \prettyref{sec:setup_obs}, the pipeline from disk model to visibilities and closure phases is presented. The results of the feasibility study can be found in \prettyref{sec:res}, and we conclude our findings in \prettyref{sec:conc}.
    
\section{Model set-up}
\label{sec:setup}
    In this section, we describe the geometrical and compositional set-up of the protoplanetary disk, the dust, and the heating sources -- in other words, the central star and the embedded companion.
    \subsection{Heating sources and brightness asymmetry}
    \label{sec:setup_asy}
        The host star of the protoplanetary disk is a black-body, low-mass, pre-main sequence (PMS) star. The stellar properties -- effective temperature ($T_{\text{eff}} = \SI{4050}{\K}$), luminosity ($L_{\star} = \SI{0.9}{\lsun}$), and distance ($d=\SI{140}{\pc}$) -- are those of the classical T~Tauri star DR~Tau \citep{brunngraeber-et-al-2016}, a prototypical PMS star. The disk asymmetries investigated here, will be located at stellar distances at least twice as large as that of the inner rim of the disk. Therefore, we restrict the parameter space to a T~Tauri star, as Herbig Ae/Be stars are significantly brighter, and thus the inner parts of the disk are hotter and brighter as well, making it even more challenging to detect these brightness asymmetries with NIR or MIR interferometers, which are mostly contrast-limited. Besides, the chosen stellar parameters ensure that the disk configurations match the flux limitations of MATISSE.
        
        In addition, a second black body is placed in the midplane of the disk, representing a companion in an early evolutionary stage, still embedded in the disk. Its effective temperature \Tc{} and luminosity \Lc{} are taken from the evolutionary tracks of planets and brown dwarfs calculated by \citet{burrows-et-al-1993,burrows-et-al-1997}. We truncate these tracks at ages higher than 20 million years, as it is expected that a protoplanetary disk vanishes within this period due to accretion, photo-evaporation, and dust evolution \citep{alexander-et-al-2014}. The total luminosity of a hotspot in a disk may be dominated by the accretion luminosity and the bright disk surrounding the companion \citep{zhu-2015,marleau-et-al-2017}. Gas shocks due to accretion may extend to several hundred planetary radii resulting in an emission which is significantly brighter than the planetary radiation alone \citep{szulagyi-mordasini-2017}. Recent simulations show that the total luminosity of high-mass planets (\SI{10}{\mjup}), including the radiation originated from the planetary accretion and the shock, can be as high as several per cent of the solar luminosity \citep{mordasini-2013,szulagyi-mordasini-2017,mordasini-marleau-molliere-2017}. An unresolved circumplanetary disk may increase the observed luminosity even further. Thus, we take luminosities of up to \SI{1}{\lsun} into account, although this luminosity must be considered as the sum of intrinsic, accretion, shock, and circumplanetary disk luminosity.
        
\begin{figure*}
            \centering
            \includegraphics[width=\hsize,height=1\vsize,keepaspectratio]{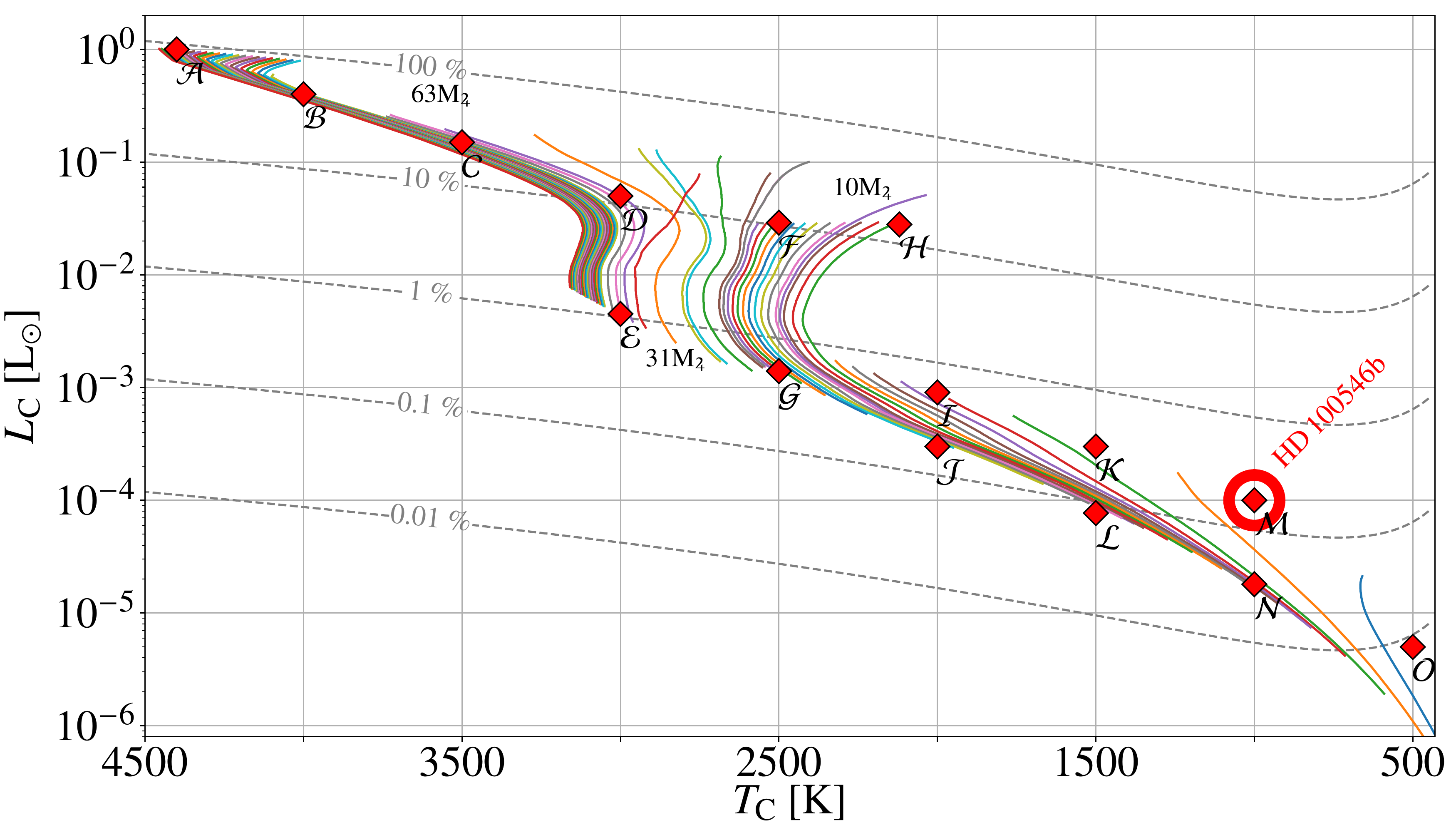}
            \caption{Parameter space of the selected effective temperatures \Tc{} and luminosities \Lc{} of the embedded companions. The coloured, solid lines show evolutionary tracks of planets and brown dwarfs as computed by \citet{burrows-et-al-1993,burrows-et-al-1997}. Sources used in our simulations are marked with red diamonds and indexed by letters (\bfcal{A} -- \bfcal{O}) in the figure. The corresponding values for \Tc{} and \Lc{} are compiled in \prettyref{tab:param_comp}. The grey, dashed contour lines indicate the flux ratio between secondary and primary source in the \textit{M} band ($\lambda = \SI{5}{\um}$). As a comparison, the derived properties of HD~100546b from \citet{quanz-et-al-2015} are also indicated.}
        \label{fig:param_space_comp}
        \end{figure*}

        In \prettyref{fig:param_space_comp}, the evolutionary tracks of those companions which fulfil the aforementioned criteria are shown. In addition to the evolutionary tracks, we have plotted contour lines in the figure, which show the NIR flux ratio of the secondary and primary source, that is, the embedded companion and the central star, respectively. For our simulations, we used a set of 15 \Lc-\Tc{} combinations, covering representative regions of the parameter space. These points are plotted on the figure. The corresponding luminosities and temperatures, \Lc{} and \Tc{}, are listed in \prettyref{tab:param_comp}. As no unambiguous (model-independent) connection between planetary mass and its luminosity and temperature can be drawn, no masses are explicitly given for the 15 considered companions. However, for the evolutionary tracks \citet{burrows-et-al-1993,burrows-et-al-1997} give masses in the range of \SIrange{1}{80}{\mjup}. These values are only true in terms of the calculation of these tracks which neglect the impact of circumplanetary disks and accretion. Each of these sources is placed at four different stellar distances $r$. The distances of \SIlist{1;2;5;10}{\au} are chosen to be close to the semi-major axes of the solar system planets (Earth, Mars, Jupiter, Saturn).
        
        \begin{table*}
            \caption{Effective temperature \Tc{} and luminosity \Lc{} of the embedded companion at different points in the parameter space (see \prettyref{fig:param_space_comp}).}
            \centering
            \begin{tabular}{l r l l r l l r l}
                \hline\hline
                \rule{0pt}{2.5ex}          & \Tc{}~[\si{\K}] & \Lc{}~[\si{\lsun}] &           & \Tc{}~[\si{\K}] & \Lc{}~[\si{\lsun}] &           & \Tc{}~[\si{\K}] & \Lc{}~[\si{\lsun}] \\[1mm]
                \hline
                \rule{0pt}{2.5ex}\bfcal{A} & 4400            & \num{1.0}                & \bfcal{F} & 2500            & \num{2.9e-2}              & \bfcal{K} & 1500            & \num{3.0e-4} \\
                                 \bfcal{B} & 4000            & \num{4.0e-1}             & \bfcal{G} & 2500            & \num{1.4e-3}              & \bfcal{L} & 1500            & \num{7.7e-5} \\
                                 \bfcal{C} & 3500            & \num{1.5e-1}             & \bfcal{H} & 2120            & \num{2.8e-2}              & \bfcal{M} & 1000            & \num{1.0e-4} \\
                                 \bfcal{D} & 3000            & \num{5.0e-2}             & \bfcal{I} & 2000            & \num{9.1e-4}              & \bfcal{N} & 1000            & \num{1.8e-5} \\
                                 \bfcal{E} & 3000            & \num{4.5e-3}             & \bfcal{J} & 2000            & \num{3.0e-4}              & \bfcal{O} & 500             & \num{5.0e-6} \\
                \hline
            \end{tabular}
            \label{tab:param_comp}
        \end{table*}
        
    \subsection{Disk}
    \label{sec:setup_disk}
        The disk is chosen to be a typical protoplanetary disk around a classical T~Tauri star. Derived from the $\alpha$-model for accretion disks from \citet{shakura-sunyaev-1973}, the density distribution is given by
        \begin{equation}
            \varrho(r,z) = \varrho_0 \left(\frac{r}{R_{0}}\right)^{-\alpha}\exp{\left[-\frac{1}{2}\left(\frac{z}{h(r)}\right)^2\right]}
        \label{eq:density_dist}
        \end{equation}
        with
        \begin{equation}
            \varrho(R_{0},0) = \varrho_0\ ,
        \end{equation}
        where $r$ and $z$ are the usual cylindrical coordinates, $\varrho_0$ the scaling factor for the total dust mass, and $h(r)$ the scale height:
        \begin{equation}
            h(r) = h_{100}\left(\frac{r}{\SI{100}{\au}}\right)^{\beta}\ .
        \label{eq:scale_height}
        \end{equation}
        The two exponents $\alpha$ and $\beta$ describe the radial density profile and disk flaring, respectively. Assuming hydrostatic equilibrium, these exponents cannot be chosen independently \citep{shakura-sunyaev-1973}:
        \begin{equation}
            \alpha = 3\left(\beta-\frac{1}{2}\right)\ .
        \label{eq:alpha_beta}
        \end{equation}
        This simple disk model has already been successfully used to fit spatially resolved and unresolved multi-wavelength observations of several protoplanetary disks \citep{burrows-et-al-1996,wolf-et-al-2003,pinte-et-al-2008,glauser-et-al-2008,ratzka-et-al-2009,sauter-wolf-2011,schegerer-et-al-2013,grafe-et-al-2013,liu-et-al-2013,brunngraeber-et-al-2016,scicluna-et-al-2016,kirchschlager-wolf-madlener-2016}. In \prettyref{tab:param_disk} the parameter values of the disk model are compiled.
        
        The dust in our set-up is a mixture of \SI{37.5}{\percent} graphite (\SI{12.5}{\percent} parallel and \SI{25}{\percent} perpendicular) and \SI{62.5}{\percent} astronomical silicate \citep{draine-lee-1984,laor-draine-1993,weingartner-draine-2001} and has particle sizes between \SIlist{5;250}{\nm}. The grain size distribution is given by the work of \citet{mathis-et-al-1977} and follows a power law, $n(s) \propto s^{-3.5}$, where $n$d$s$ is the number of particles in the radius interval $[s,s+\text{d}s]$.
        
        \begin{table}
            \caption{Parameters of the disk model and the central star.}
            \centering
            \begin{tabular}{l l}
                \hline\hline
                \rule{0pt}{2.5ex}Disk or stellar parameter        & Value \\[1mm]
                \hline
                \rule{0pt}{2.5ex}$L_{\star}$~[\si{\lsun}]      & 0.9    \\
                $T_{\text{eff}}$~[\si{\K}]                     & 4050   \\
                $\log{\nicefrac{M_{\text{dust}}}{\si{\msun}}}$ & -7 .. -3 ($\Delta$ = 0.5) \\
                $R_{\text{in}}$~[\si{\au}]                     & 0.5    \\
                $R_{\text{out}}$~[\si{\au}]                    & 100    \\
                $\beta$                                        & 1.1    \\
                $h_{\text{100}}$~[\si{\au}]                    & 14     \\
                Inclination $\Theta$~[\si{\degree}]            & 0      \\
                Distance $d$~[\si{\pc}]                        & 140    \\
                Stellar separation of companion $r$~[\si{\au}] & \{1, 2, 5, 10\} \\
                \hline
            \end{tabular}
            \label{tab:param_disk}
        \end{table}
    
    \subsection{Deriving the observables}
    \label{sec:setup_obs}
\begin{figure}
            \centering
            \includegraphics[width=\columnwidth,height=0.472\textheight,keepaspectratio]{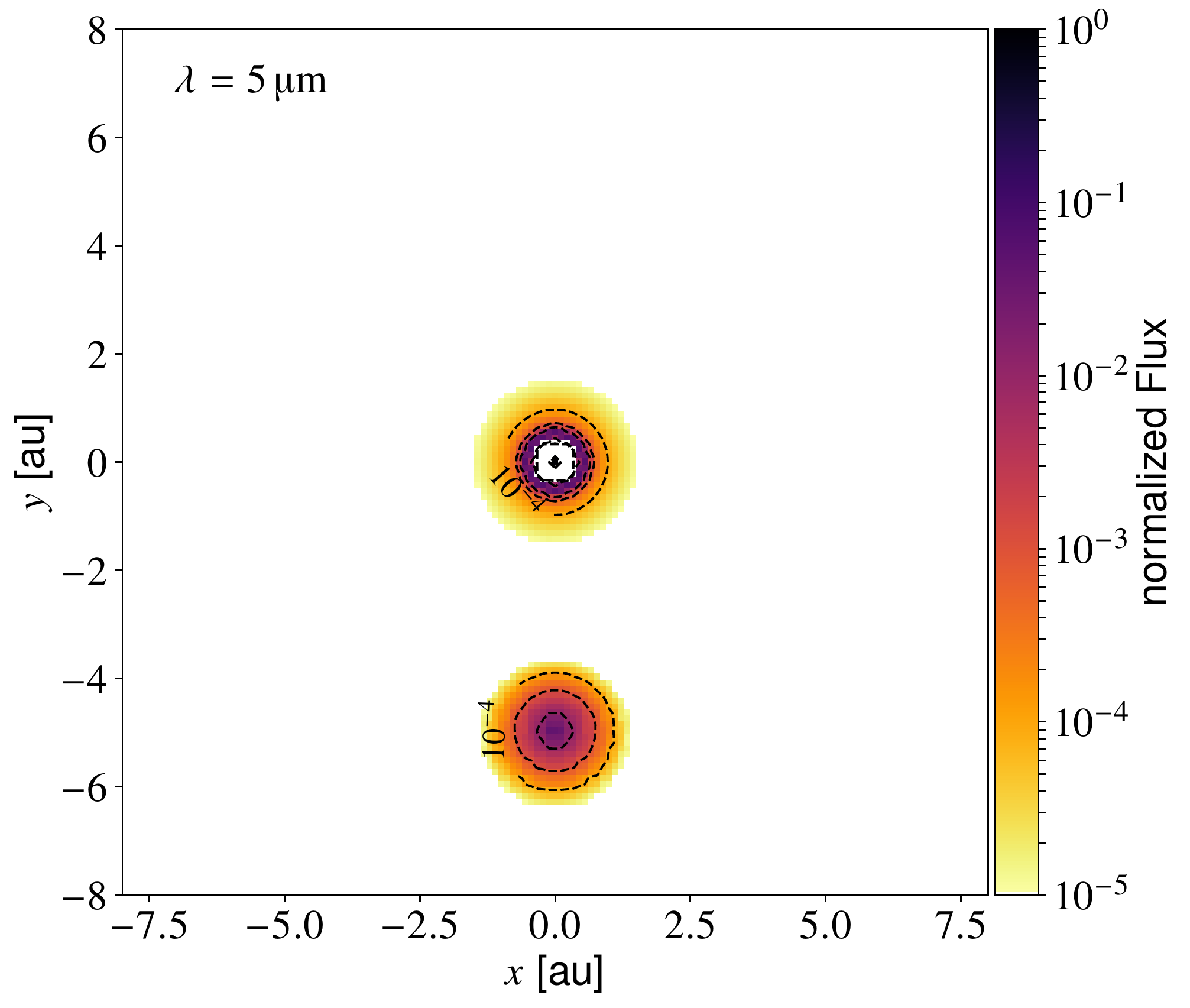}\\
            \includegraphics[width=\columnwidth,height=0.472\textheight,keepaspectratio]{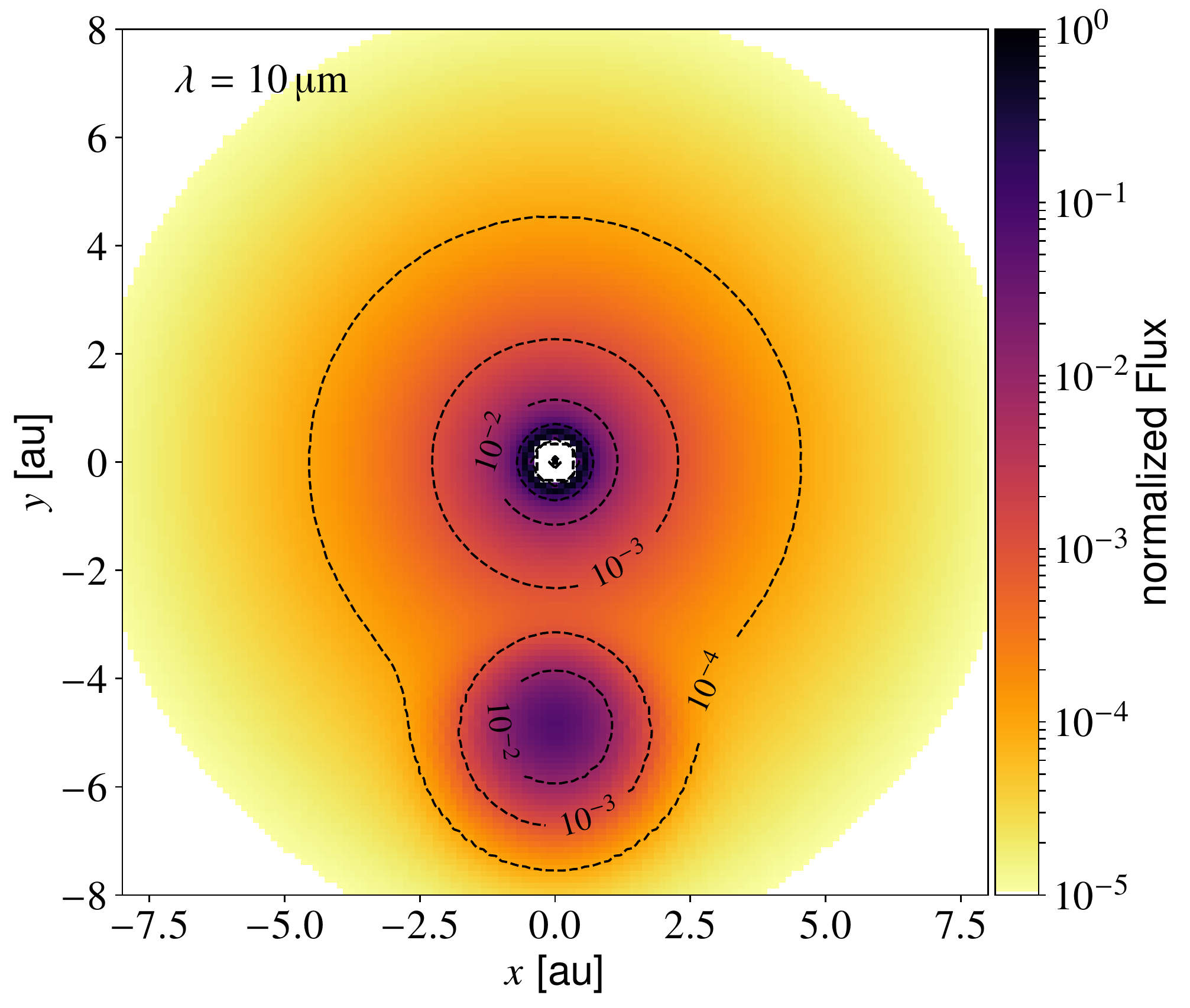}
            \caption{Exemplary combined re-emission and scattered light map of our star-companion-disk model (dust mass of \SI{1e-5}{\msun}, object \bfcal{C} in \prettyref{tab:param_comp}, $r = \SI{5}{\au}$). \textit{Top}: $\lambda = \SI{5}{\um}$ (\textit{M} band). \textit{Bottom}: $\lambda = \SI{10}{\um}$ (\textit{N} band).}
            \label{fig:flux_map}
        \end{figure}

        To calculate the temperature distribution and thermal re-emission of the dust and the scattered stellar light, we use the Monte-Carlo radiative transfer code \textsc{Mol3d} \citep{ober-et-al-2015}. In addition to the central star, we have included the companion as a second source in the midplane of the disk. Thus, the dust is heated by both sources, generating a non-rotationally symmetric brightness distribution; see \prettyref{fig:flux_map}.
        
        Based on the simulated wavelength-dependent intensity maps, visibilities and closure phases are calculated via FFT. These maps are zero-padded to increase the spatial frequency resolution -- with a minimum resolution of \SI{0.5}{\Mlambda} -- and decrease the computation time. We use 29 wavelengths in total, linearly distributed over the three bands \textit{L} (\SIrange{2.78}{4.05}{\um}, $R=4$), \textit{M} (\SIrange{4.52}{4.99}{\um}, $R=1.5$) and \textit{N} (\SIrange{8.01}{13.07}{\um}, $R=6$--$10$). The spectral resolution $R$ of our set of maps is lower than what MATISSE will be capable of even in low-res mode ($R=30$). However, a higher spectral resolution would yield no additional information as the absorption and scattering cross sections of the underlying dust model show no narrow spectral features.
        
        The direction from central star to the companion is described by the angle $\Phi$ which gives the on-sky orientation measured from north (N) to east (E). In a similar way we defined the position angle ($P\!A$) of the simulated observations, that is, the projected baseline orientation, from N to E. Thus, for the observing angle $\Psi \coloneqq |P\!A - \Phi| = \SI{0}{\degree}$, the projected baseline is parallel to the connection line star - companion, and perpendicular for $\Psi = \SI{90}{\degree}$.
        
        To investigate the feasibility to detect the asymmetry we calculated the visibilities and closure phases for the UT quadruplet at hour angles ($H\!A$) between \SIlist[list-units=repeat,explicit-sign=-,retain-explicit-plus]{-3;+3}{\hour} with a five-minute resolution -- thus the object is at least \SI{30}{\degree} above the horizon for all hour angles -- , for a uniformly distributed observing angle $\Psi$ between \SIlist[list-units=repeat,]{0;90}{\degree} with a resolution of \SI{1}{\degree}, and a target declination of $\delta$~=~+17:00:00 which is within the \textit{Taurus-Aurigae} star forming region. This was done for all companion-disk combinations and for the undisturbed reference disks.
        
        The parameter space covers 549 disks $\times$ (six visibilities + three closure phases) $\times$ 29 wavelengths $\times$ 73 hour angles $\times$ 91 observing angles $\Psi$. To analyse the feasibility to observe the brightness asymmetry, the visibility and closure phase information were post-processed as follows:
        \begin{enumerate}
            \item We calculated the differences of visibilities and closure phases between asymmetric disk and reference disk.
            \item We calculated the mean value of these differences for all observing angles $\Psi$ as the orientation of the companion is unknown prior to the observation.
            \item We found the maximum for all $H\!A$ values, meaning that we assumed that we were observing at the $H\!A$ with the best ($u$,$v$)-coverage.
            \item We found the maximum for all six baselines and wavelengths per band. This approach is justified by the fact that all information of these quantities are gathered simultaneously during one observation.
        \end{enumerate}
        
        In addition, the predicted flux limitations of MATISSE (\SI{0.05}{\jy} for \textit{L} and \SI{0.12}{\jy} for \textit{N} band with external fringe tracker; see \citealp{matter-et-al-2016}) were considered.
        
    \subsection{Uncertainties}
    \label{sec:uncertain}
        Whether disk asymmetries will be detected or not, strongly depends on the uncertainties of the visibility and closure phase measurements. As MATISSE is not commissioned yet, no reliable on-sky data are available. However, during the test phase for the Preliminary Acceptance in Europe (PAE) of MATISSE, it was found that the visibility and closure phase measurements are very stable. In lab experiments, visibility uncertainties of 0.005 (\textit{L} band), 0.004 (\textit{M} band) and 0.025 (\textit{N} band), and closure phase uncertainties of \SI{18}{\arcminute} (\textit{L} and \textit{N} band) and \SI{9}{\arcminute} (\textit{M} band) were derived (A.~Matter and S.~Lagarde, private communication). These values correspond to the root mean square of the fluctuations over a time span of four hours for the \textit{L} and \textit{M} bands, and even three days for the \textit{N} band, which is much longer than the expected 30 minutes between target and calibrator observations. As these values represent only the intrinsic, instrumental errors and do not account for photon noise of dim targets, thermal background, calibration errors or fundamental noise, these values only represent ideal limits as compared to the yet unknown on-sky performance of the instrument. Our analysis therefore focusses on the necessary performance of MATISSE instead -- or any future interferometer in the same wavelength regime -- to detect disk asymmetries based on embedded companions. We used the 2-$\sigma$ levels of the obtained \textit{M} band uncertainties of the visibility and closure phase lab experiments, that is \SI{1}{\percent} and \SI{18}{\arcminute} ($\approx\num{5}$\,mrad), as the lower detection boundaries for our further analysis.
        
\section{Results}
\label{sec:res}
\begin{figure*}
        \centering
        \includegraphics[width=0.9\hsize,height=0.9\textheight,keepaspectratio]{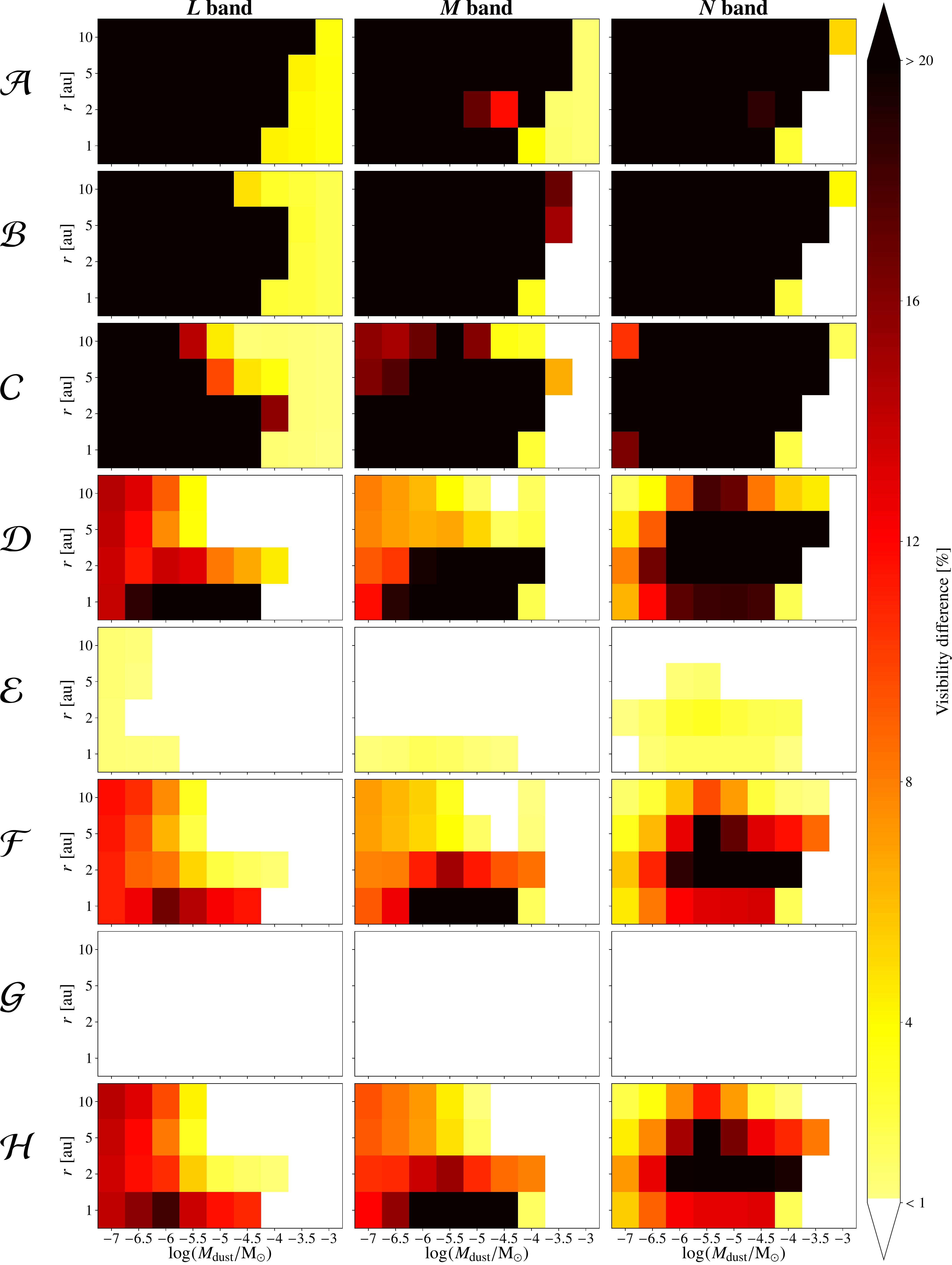}
        \caption{Maximum visibility difference of disturbed to undisturbed disk (see \prettyref{sec:setup_obs}) in the \textit{L} (\textit{left}), \textit{M} (\textit{middle}) and \textit{N} (\textit{right}) bands for eight out of the 15 disk-companion configurations that exceed at least \num{0.01} or \num{5}\,mrad, respectively. A detection above these limits of the companions \bfcal{I} to \bfcal{O} is not feasible and the corresponding plots are omitted.}
        \label{fig:detect_vis}
    \end{figure*}

\begin{figure*}
        \centering
        \includegraphics[width=0.9\hsize,height=0.9\textheight,keepaspectratio]{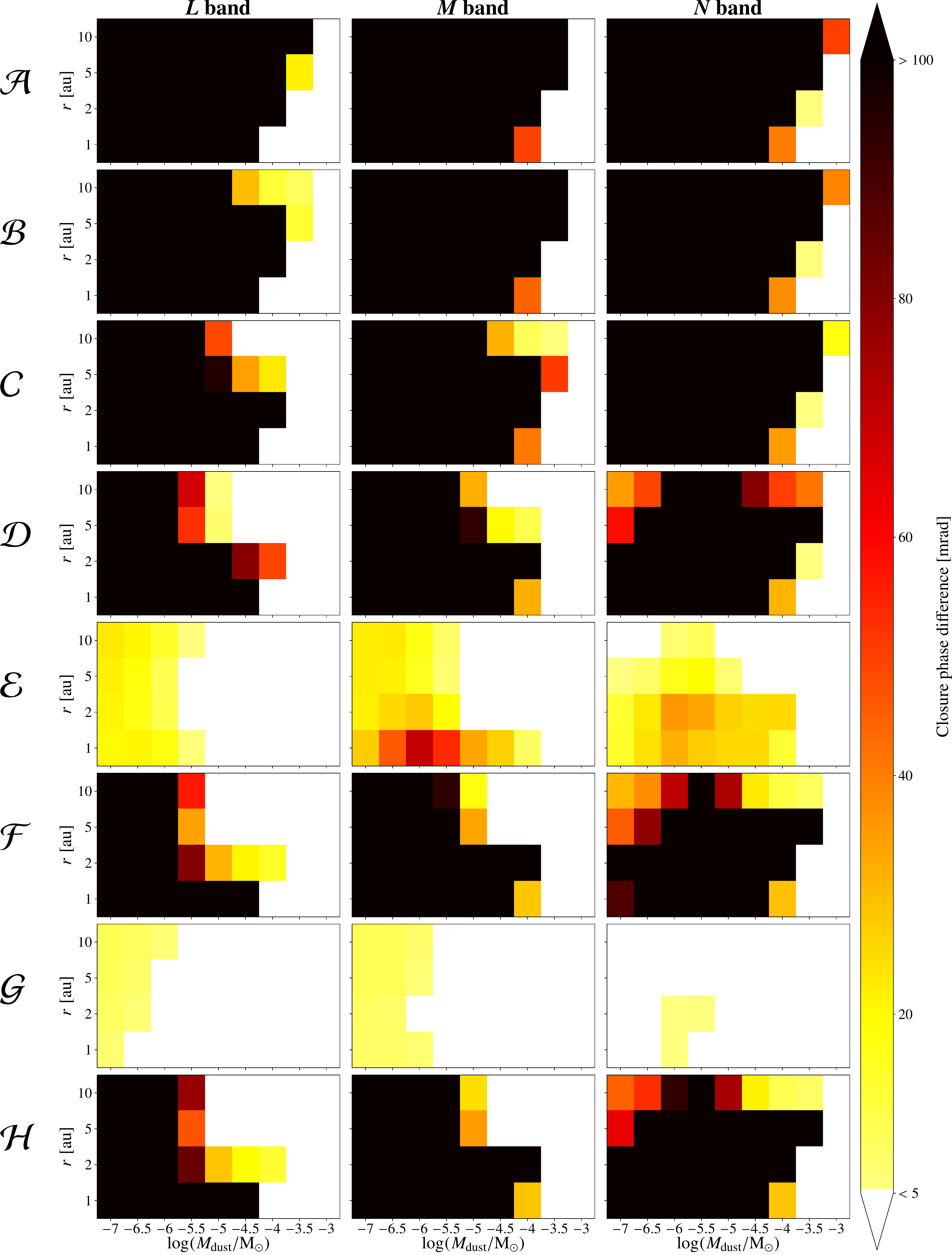}
        \caption{Same as \prettyref{fig:detect_vis} but for the closure phase measurements.}
        \label{fig:detect_clopha}
    \end{figure*}

    In this section we present the results of our radiative transfer simulations of a protoplanetary disk hosting embedded planets with the subsequent analysis of the brightness distribution in terms of a feasibility study for MATISSE/VLTI.
    The following analysis is based on the post-processed visibilities and closure phases as described in \prettyref{sec:setup_obs}. In Figs.~\ref{fig:detect_vis} and \ref{fig:detect_clopha}, we show the visibility and closure phase differences for those companions which exceed at least \SI{1}{\percent} or \num{5}\,mrad, respectively. The influence of radial distance, disk mass, companion properties and wavelength bands are described in the following subsections.
    
    \afterpage{\clearpage}  
    
    \subsection{Stellar separation}
    \label{sec:stellar_sep}
\begin{figure}
            \centering
            \resizebox*{\hsize}{!}{\includegraphics{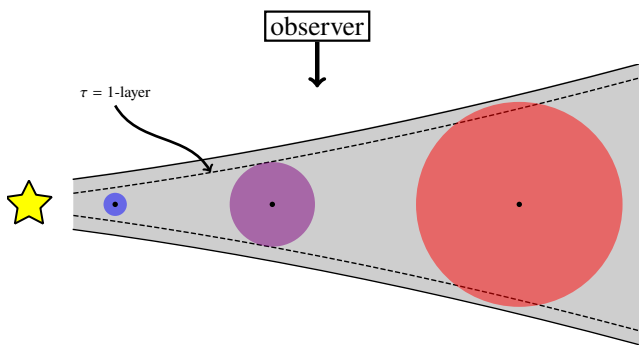}}
            \caption{Illustration of the heated region in the vicinity of the companion for three different stellar separations. The dashed line corresponds to the $\tau=1$-layer as seen from the observer who is located perpendicular to the disk midplane, i.e. approximately the region where the detected flux is emitted. The companions are shown as black points on the midplane of the disk. The heated regions due to the radiation of the companion are shown as coloured circles, where blue is the hottest region, red the coolest and purple is in between. As the optical depth decreases for increasing distance, the region heated by the embedded companion becomes larger. Therefore the photon energy is distributed over more dust particles and the temperature increase is less than for smaller stellar separations. At the same time, the region which is significantly affected by the heating of the embedded companion approaches the $\tau=1$-layer and eventually exceeds this layer, see \prettyref{fig:deltaT_radprof}. The thermal radiation of that hot dust may then leave the disk towards the observer less weakened compared to regions below the $\tau=1$-layer.}
            \label{fig:scheme}
        \end{figure}

\begin{figure}
            \centering
            \includegraphics[width=0.49\hsize,height=\vsize,keepaspectratio]{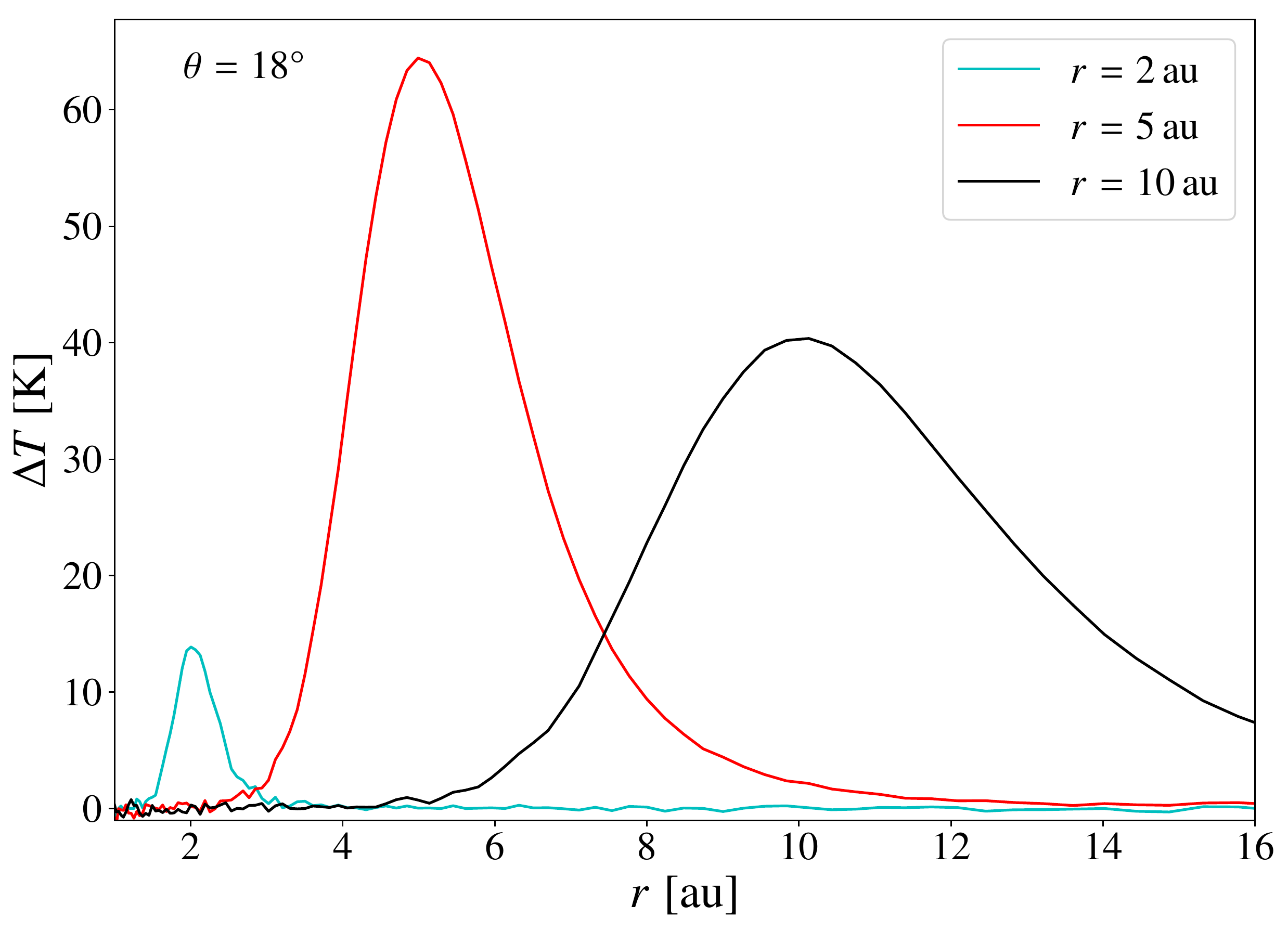}
            \includegraphics[width=0.49\hsize,height=\vsize,keepaspectratio]{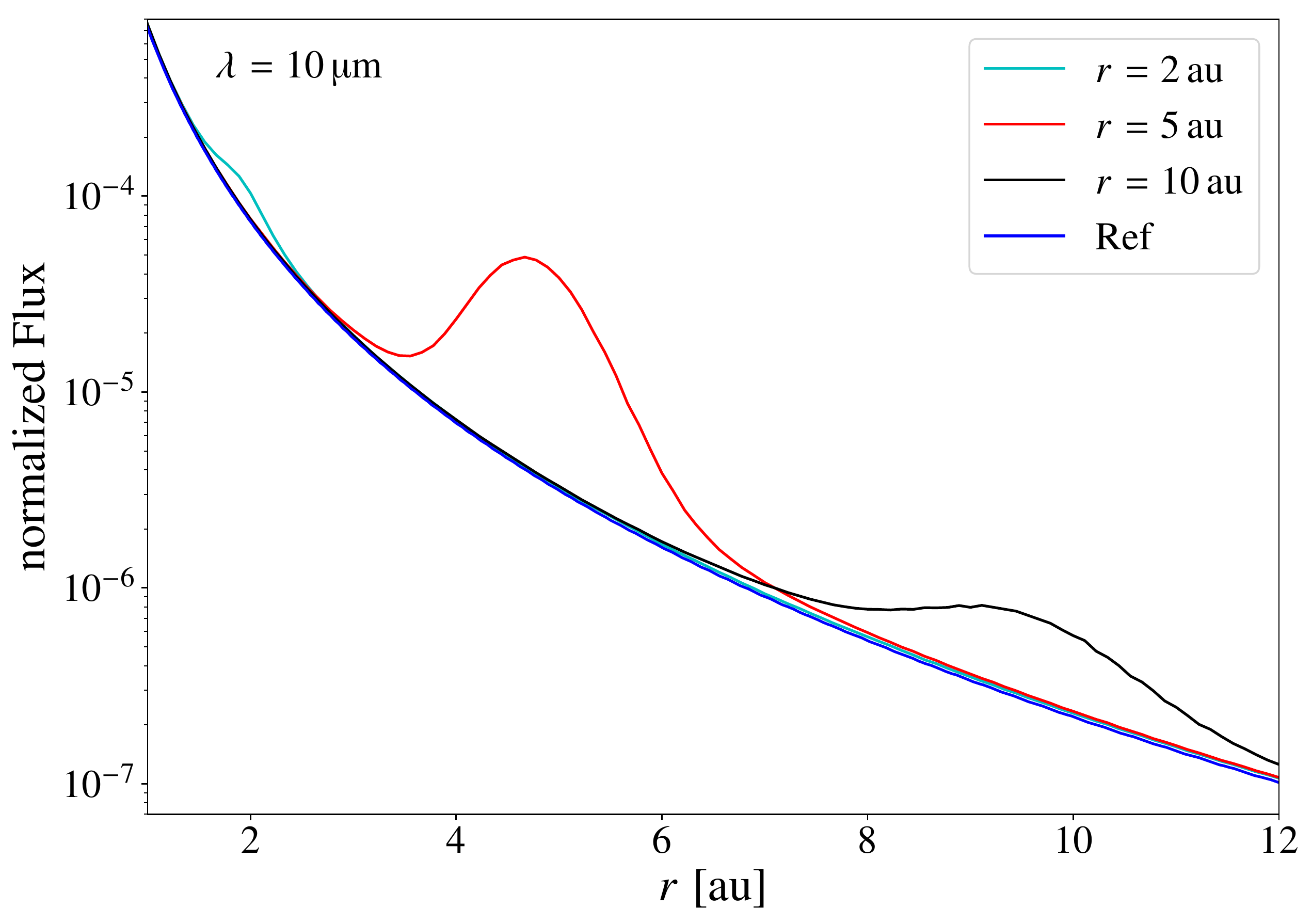}
            \caption{Temperature difference of disturbed and undisturbed disk at \SI{18}{\degree} above the midplane (\textit{left}) and radial profiles of the intensity maps at \SI{10}{\um} (\textit{N} band) (\textit{right}) of a massive disk ($\log{\nicefrac{M_{\text{dust}}}{\si{\msun}}} = -3.5$) with companion \bfcal{F} as a second heating source at \SIlist{2;5;10}{\au}. For larger stellar separations, the heated area becomes larger but the temperature increase is lower than for smaller distances, and thus the flux excess above the reference disk is lower. As seen from the observer, \SI{18}{\degree} above the midplane is close to the $\tau=1$-layer.}
            \label{fig:deltaT_radprof}
        \end{figure}

        Due to the additional heating of the embedded companion the temperature of the dust in the vicinity of that source is increased as compared to the reference case. The dust number density decreases for larger distances $r$ and as a result, the optical depth per unit length decreases as well. The mean free path length of photons released from the companion increases and the photon energy is distributed over a larger spatial region. Thus, the temperature offset in the environment of the embedded source in comparison to the undisturbed disk decreases further away from the star. Besides, in the NIR to MIR wavelength regime, the observed flux originates in the upper layers of the disk. The optical depth from the disk midplane to these layers decreases as well for increasing distance. Therefore, the regions heated by the companion may well reached into the upper disk layers for larger distances. This effect is schematically depicted in \prettyref{fig:scheme}. In the left panel of \prettyref{fig:deltaT_radprof}, the temperature difference at the $\tau=1$-layer (about \SI{18}{\degree} above the midplane for the \textit{N} band) between disturbed and undisturbed high-mass disks, $\log{\nicefrac{M_{\text{dust}}}{\si{\msun}}} = -3.5$, is exemplary shown for distances of \SIlist{2;5;10}{\au} and the companion \bfcal{F}. At \SI{18}{\degree} above the midplane, the temperature offset for a distance of \SI{2}{\au} is much smaller than for $r=\SI{5}{\au}$. This can be explained with the decreasing vertical optical depth. Further, the offset for $r=\SI{10}{\au}$ decreases again because of the larger region over which the photon energy is distributed. This has direct consequences on the observed flux. Radial profiles of the flux maps are shown in the right panel of \prettyref{fig:deltaT_radprof}. In terms of the detection significance, this effect can be seen very clearly in Figs.~\ref{fig:detect_vis} and \ref{fig:detect_clopha} for the sources \bfcal{D}, \bfcal{F} and \bfcal{H}.
        
\begin{figure}
            \centering
            \includegraphics[width=\hsize,height=\vsize,keepaspectratio]{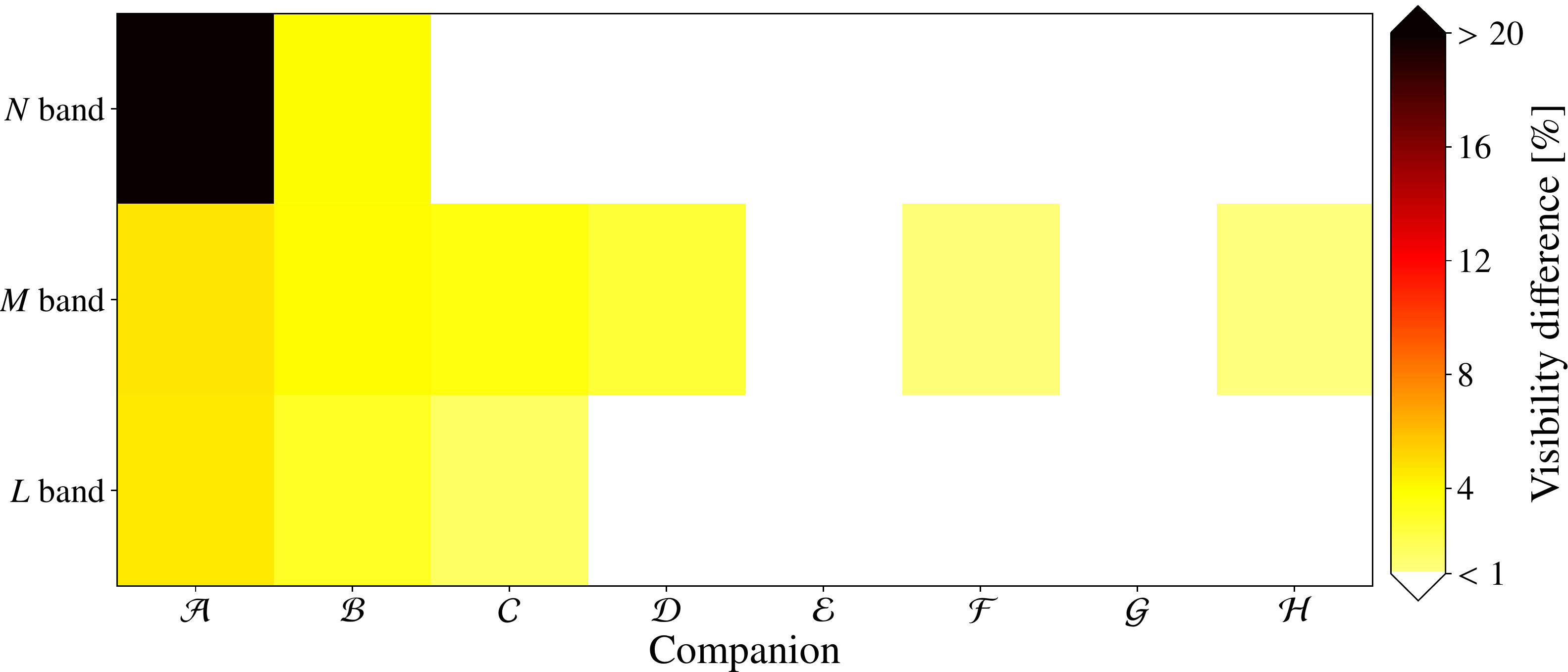}
            \caption{Same as \prettyref{fig:detect_vis} but for a fixed radial distance of $r=\SI{50}{\au}$ to investigate the feasibility to trace protoplanet candidates detected via direct imaging with MATISSE.}
            \label{fig:detect_50au}
        \end{figure}

        In addition, we also performed simulations that had a stellar separation of $r=\SI{50}{\au}$ and a dust mass of $M_{\text{dust}} = \SI{1e-4}{\msun}$ in order to investigate the feasibility to trace known protoplanet candidates -- such as HD~100546b -- with MATISSE. All visibility differences between disturbed and undisturbed disk are less than \SI{4}{\percent} except for the very high-luminous companion \bfcal{A} in the \textit{N} band, see \prettyref{fig:detect_50au}. Significant closure phases can also only be expected for the \textit{N} band and the two most luminous companions.
        
    \subsection{Planetary radiation}
        The luminosity of the companion is the decisive factor which determines the feasibility to observe the companion, while its temperature is of minor importance only within the given parameter space. The objects \bfcal{E} and \bfcal{G} are much harder to detect than the objects \bfcal{D}, \bfcal{F} and \bfcal{H} although they are in the same temperature range. Especially the difference between object \bfcal{E} and the nearly \SI{1000}{\K} cooler but nearly ten times more luminous object \bfcal{H} is significant. The dust absorption cross section $C_{\text{abs}}$ at the wavelength of maximum radiation of the companion varies only for a factor of about two between $\Tc=\SI{3000}{\K}$ (\bfcal{E}) and $\Tc=\SI{2000}{\K}$ (\bfcal{H}). More available energy to heat the dust, that is, a higher planetary luminosity, corresponds to a brighter -- and thus easier to observe -- asymmetry.
    
    \subsection{Disk mass}
        For most of the disk-companion configurations, we find that the disk mass does not change the feasibility to detect the asymmetries if $\log{\nicefrac{M_{\text{dust}}}{\si{\msun}}} \le -4$. There are some exceptions though:
        \begin{itemize}
            \item For distances of \SIlist{5;10}{\au}, the highest disk mass for which the asymmetry can be detected is increased and decreased, respectively, due to the effects mentioned in \prettyref{sec:stellar_sep} in other words, the local heating is possible over larger regions and thus the smaller temperature excesses for $r=\SI{10}{\au}$.
            \item Because of the increase of the optical depth for high-mass disks with $\log{\nicefrac{M_{\text{dust}}}{\si{\msun}}} > -4$, only the brightest companions can be detected via their visibilities. However, this result may change if gaps were included in the disk-companion model. The presence of a gap will increase the contrast between disk and planet making it more feasible to detect dimmer companions \citep{wolf-dangelo-2005}.
        \end{itemize}
        
\begin{figure}
            \centering
            \includegraphics[width=0.49\hsize,height=\vsize,keepaspectratio]{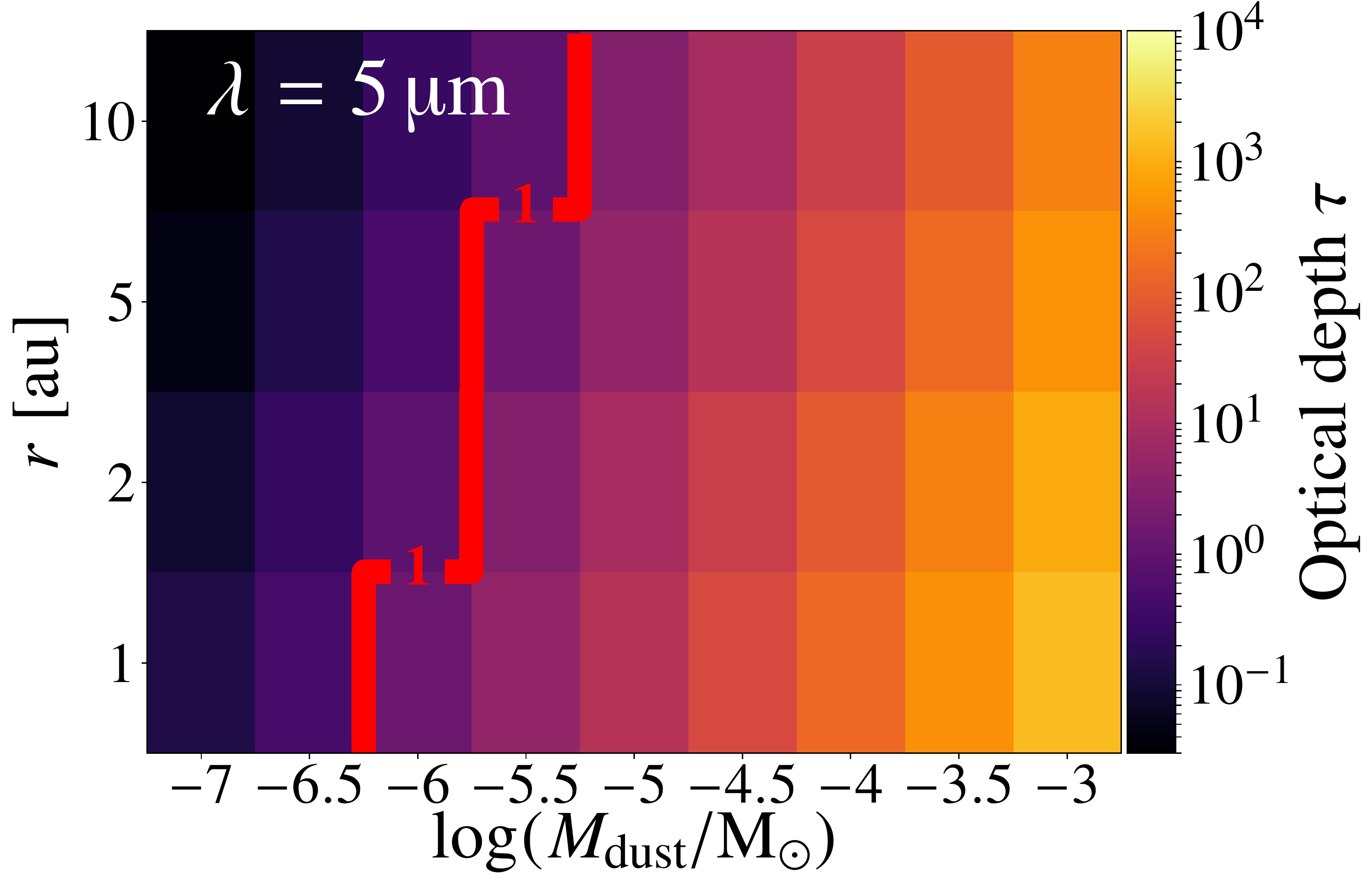}
            \includegraphics[width=0.49\hsize,height=\vsize,keepaspectratio]{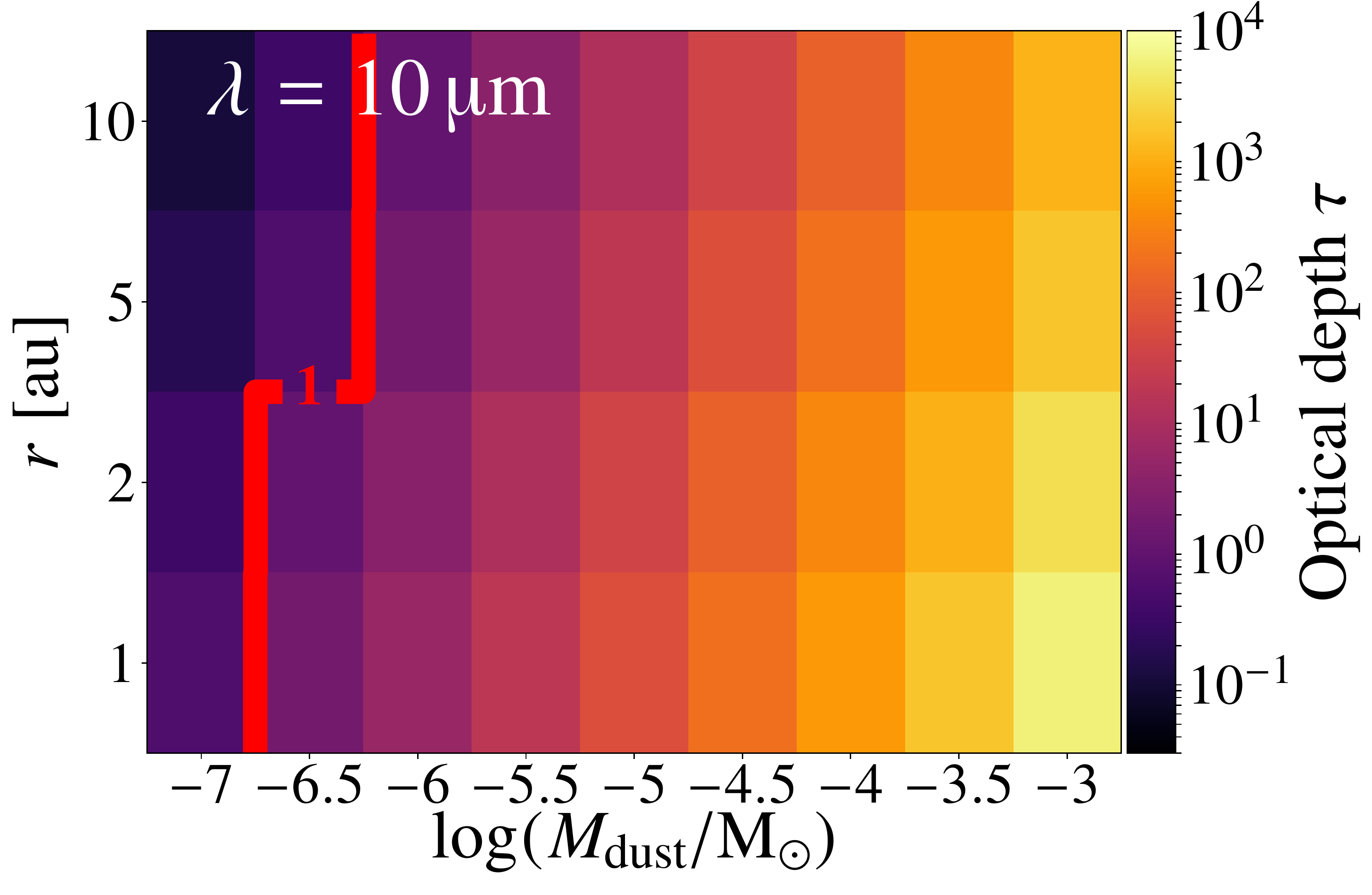}
            \caption{Vertical optical depth from the disk midplane towards the observer for a wavelength $\lambda$ of \SI{5}{\um} (\textit{left}) and \SI{10}{\um} (\textit{right}). The transition from the optical thick ($\tau>1$) to the optical thin case ($\tau<1$) is marked by a red contour line.}
            \label{fig:opt_depth}
        \end{figure}

        For an overview of the vertical optical depth --  measured from the disk midplane to the observer -- of the considered disks see \prettyref{fig:opt_depth}. No clear evidence can be found in Figs.~\ref{fig:detect_vis} or \ref{fig:detect_clopha} that the transition from optically thin to optically thick may affect the feasibility to detect the brightness asymmetries caused by the companions.
        
    \subsection{Visibility and closure phase}
        Considering the applied lower limits, the closure phase measurements are suitable to detect a larger fraction of the investigated disk-companion parameter space with a higher SNR than the visibility measurements. This is especially true for the low-luminous companions \bfcal{E}, \bfcal{G} and \bfcal{H}. The phase information of a Fourier transformation is in general more sensitive to radial asymmetries in the original image than the visibility. Only for very high-mass disks and high-luminous companions, does the visibility provide a higher feasibility to detect the hot spots.
        
    \subsection{Wavelength}
        For most of the considered parameter space, the three wavelength bands (\textit{L}, \textit{M}, \textit{N}) provide similar visibility and closure phase differences compared to the reference cases. However, only in the \textit{L} band the high-luminous companions \bfcal{A}, \bfcal{B} and \bfcal{C} can be detected if embedded in disks with the highest dust masses. Because of the high temperature, the companions radiation, and thus the scattered light, peaks at short wavelengths and the ability to detect the scattered light decreases with increasing wavelength. However, for the companions \bfcal{D} and lower, the feasibility to detect the asymmetries increase for increasing wavelength because the thermal re-emission of the surrounding dust predominates its appearance. The dust temperature in the relevant disk regions is in the range of some hundred Kelvin and thus the re-emission peaks in the MIR wavelength range rather than in the NIR. In addition, the optical depth decreases for larger wavelengths and the radiation can leave the disk less weakened.
    
\section{Conclusions}
\label{sec:conc}
    We analysed the feasibility to detect protoplanetary disks hosting young, embedded protoplanets, generating an asymmetric brightness distribution, with the upcoming second generation VLTI instrument MATISSE. The following results were obtained:
    \begin{itemize}
        \item The visibility and closure phase differences between the asymmetric disk and undisturbed reference disk exceed at least \SI{1}{\percent} or \num{5}\,mrad, respectively, for 8 of the 15 companions considered. The corresponding minimum flux ratio of companion and central star can be as small as \SIrange{0.5}{0.6}{\percent}. This ratio is in the order of what is expected for mid- to high-mass jovian planets found in latest planet formation simulations (e.g. \citealt{mordasini-2013,szulagyi-mordasini-2017,mordasini-marleau-molliere-2017}).
        \item Within the assumed measurement accuracies closure phase measurements were found to be more suitable to detect the companions compared to visibility measurements. This is because phase measurements are in general much more sensitive to radial asymmetries in the brightness distribution.
        \item The stellar separation of the asymmetry should be in the intermediate region of $r\approx\SIrange{2}{5}{\au}$. For smaller separations the optical depth increases rapidly making it more difficult for the emitted photons to reach upper disk layers. If the companion is at distances of $r\approx\SI{10}{\au}$ from the star, the photons cover much larger distances which results in a more expand hotspot. Therefore, the energy input per area, and thus the resulting temperature, is reduced and the hotspot is less pronounced.
        \item For very high-mass disks the detectability decreases rapidly even for the brightest planets. The highest detecting probability can be expected for disks with dust masses of $M_{\text{dust}}\le \SI{1e-4}{\msun}$.
    \end{itemize}
    
    These results may change if the fundamental disk-asymmetry-model is different. Massive jovian planets would likely interfere with the gas and dust distribution creating spiral arms, gaps, a circumplanetary disk or other large scale structures. These structures may help to draw conclusions on whether a planet is present in the disk, although planets are not the only reason for such signatures. Furthermore, if the planet has already cleared a gap, the detectability may increase because of a higher contrast between the gap and the planet with its protoplanetary disk \citep{wolf-dangelo-2005}. Besides, hot dust located in a limited region would cause the scale height of the disk to locally increase due to the hydrostatic equilibrium, increasing the scattered stellar light on that location.
    
    Whereas high-contrast NIR imagers such as SPHERE/VLT or GPI/Gemini are capable of detecting the aforementioned large-scale structures created by planet-disk interaction, MATISSE might enable the observer to confirm the presence of a planet and its location by directly detecting the point-like emission of that planet or the nearby hot dust. The important advantage of MATISSE over single-telescope imagers is the high spatial resolution of several to some ten milli arcseconds over a large wavelength range. As a comparison, the known protoplanet candidates HD~100546b -- detected via direct imaging  -- and LkCa~15b -- detected by the LBT interferometer -- have stellar separations of about 500 and 80\,mas, respectively.
    
    Although the Atacama Large Millimeter/submillimeter Array (ALMA) provides very long baselines of several kilometres, the spatial resolution is still a factor of between three and ten lower than a NIR/MIR interferometer with baselines of up to \SI{130}{\m}. However, \citet{wolf-dangelo-2005} already showed that planets may be detectable with ALMA through their circumplanetary accretion region, and recent studies also suggest that ALMA is capable of detecting these circumplanetary disks at regions of about \SI{5}{\au} from the star for a wavelength of $\lambda=\SI{400}{\um}$ \citep{szulagyi-et-al-2017}, and for stellar separations of \SIrange{20}{50}{\au} for $\lambda=\SI{870}{\um}$ \citep{zhu-andrews-isella-2017,szulagyi-et-al-2017}. For a separation of \SI{50}{\au} such as for HD~100546b detections may also be achievable using high-resolution line observations of \element[][12]{CO} and \element[][13]{CO} \citep{perez-et-al-2015}.

\begin{acknowledgement}
    This research was funded through the DFG grants WO 857/13-1 and WO 857/17-1. The authors are very greatful for the enlightening correspondence with A. Matter. R.B. thanks Robert Brauer for his help and ideas during their various discussions.
\end{acknowledgement}




\bibliographystyle{aa}
\bibliography{lit}


\end{document}